\documentclass{article}

\usepackage{amsfonts,amssymb}

\newcommand{\slsh}[1]
{
#1 \!\!\!\slash
}

\def\G{\Gamma}
\def\GG{{\rm I}\!\Gamma}

\begin{document}

\rightline{IFUM-824-FT}

\vskip 2.4 truecm
\Large
\bf
\centerline{Slavnov-Taylor parameterization}
\par \centerline{of Yang-Mills theory with massive fermions}
\par \centerline{in the presence of singlet axial-vector currents} 
\normalsize \rm

\vskip 1.4 truecm
\large
\centerline{Andrea Quadri
\footnote{E-mail address: {\tt andrea.quadri@mi.infn.it}}}

\vskip 0.3 truecm
\normalsize
\centerline{Dip. di Fisica, Universit\`a di Milano, 
via Celoria 16, 20133 Milano, Italy } 
\centerline{and I.N.F.N., sezione di Milano} 

\vskip 0.7  truecm
\normalsize
\bf
\centerline{Abstract}

\rm

\begin{quotation}
We study the all-order restoration of the Slavnov-Taylor
(ST) identities for Yang-Mills theory with massive fermions in the
presence of singlet axial-vector currents. 
By making use of the ST parameterization
of the symmetric quantum effective action 
a natural set of normalization conditions is derived
allowing to reduce the algebraic complexity 
of higher orders ST identities up to a homogeneous linear problem.
Explicit formulas for the action-like part of the symmetric
vertex functional are given to all orders in the loop expansion.
\end{quotation}

\newpage

\vskip 0.8 truecm

\section{Introduction}
\label{sec:intr}
In the BRST quantization \cite{Becchi:1974md}-\cite{Piguet:1995er}
of gauge theories the requirement of
Physical Unitarity is guaranteed by the fulfillment of the
Slavnov-Taylor (ST) identities \cite{Slavnov:1972fg,Taylor:1971ff}.

Even in the absence of anomalies the ST identities may be spoiled by
the regularization procedure used to remove UV divergences in the
Feynman amplitudes. 
For theories involving the $\gamma^5$ matrix and the completely
antisymmetric tensor dimensional regularization is not an invariant
regularization scheme.
An invariant regularization for anomaly-free chiral gauge theories
like the Standard Model and the SO(10) chiral theory has been 
introduced in \cite{Frolov:1992ck}-\cite{Bakeyev:1996is}.
On the other hand, in the context of supersymmetric theories
no all-order invariant regularization scheme is known.

By exploiting cohomological techniques  it can be proven 
in a regulari\-zation-independent way that,
if the model is anomaly-free, an order-by-order choice 
of non-invariant counterterms exists allowing
to recursively fulfill the ST identities.
Along these lines the Standard Model (SM) has been dealt with in  
\cite{Grassi:1995wr,Grassi:1999nb,Kraus:1997bi} and
the Minimal Supersymmetric Standard Model in \cite{Hollik:2002mv}.

On the other hand, there is no completely satisfactory answer to the
 question of how to find the explicit form 
of the non-invariant symmetry-restoring counterterms.
Modified subtraction schemes have been considered 
e.g. in \cite{Aoyama:1980yw}-\cite{Stockinger:2005gx}.
Methods based on the computation of the ST breaking terms
have been applied in \cite{Ferrari:1998jy} and
\cite{Martin:1999cc}. 
Within the framework of Algebraic Renormalization \cite{Piguet:1995er}
the direct imposition of the ST identities after expansion of the
symmetric vertex functional on Lorentz-invariant monomials, compatible
with all unbroken symmetris of the model,
has been proposed in \cite{Ferrari:1999nj} 
and applied to some specific SM examples
in \cite{Grassi:1999tp}-\cite{Grassi:2001kw}.
In \cite{Grassi:2001zz} simplifications coming from the use of
the background field method have also been discussed.
Proper subsets of relations between Green functions, stemming from
the ST identities, have been considered
for several supersymmetric models in
\cite{Hollik:1999xh}-\cite{Fischer:2003cb}.

In the absence of IR problems  
an alternative proposal  was suggested in \cite{Quadri:2003ui}.
The finite terms
of the action-like part of $\GG$ are constructively parameterized
by making use of the elements of the cohomology 
of the classical linearized ST operator ${\cal S}_0$ without power-counting 
restrictions (ST invariants) order by order in the loop
expansion.
This in turn allows for a  systematic explicit derivation of 
the non-invariant counterterms  recursively needed in order 
to restore the ST identities.

The purpose of the present paper is to apply such a technique to the 
study of Yang-Mills theory
with massive fermions in the presence of singlet axial-vector currents.

This model is directly related to the QCD sector of the Standard Model.
There the singlet axial-vector currents characterize the couplings of the 
neutral vector bosons to the quarks. Two-loop QCD corrections
to the heavy quark form factors have been recently discussed
in the framework of dimensional regularization 
in \cite{Bernreuther:2004ih, Bernreuther:2004th}.

Here we choose to work with the non-invariant BPHZL procedure
in order to regularize the Feynman amplitudes.
The ST parameterization is constructed for the model at hand
and used to study the properties of the action-like part of the symmetric
quantum effective action, i.e. the projection of the 1-PI vertex functional
onto the sector spanned by all possible Lorentz-invariant
monomials of dimension $\leq 4$ in the fields, the antifields,
the external sources and their derivatives, compatible
with all unbroken symmetries of the theory.

\noindent
When cohomologically non-trivial masses  (unlike IR regulators) are present,
the finite $n$-th order 
non-symmetric action-like terms depend on superficially convergent $n$-th order
Feynman amplitudes. This property has been pointed out for the Abelian
Higgs-Kibble model in \cite{Ferrari:1999nj}. 
By using the ST parameterization, 
the underlying geometrical structure is investigated and clarified here
in a full non-Abelian setting. 

In \cite{Ferrari:1999nj} it was also 
noted that, due to the presence of physical mass parameters,
 a number of consistency conditions
involving superficially convergent Feynman amplitudes appears.
The origin of these relations can be traced back to the Wess-Zumino consistency
conditions for the ST breaking terms \cite{Ferrari:1999nj}. 
This becomes particularly clear when ST invariants are used in order to parameterize the symmetric
quantum effective action. This issue is illustrated on the non-trivial
set of consistency conditions arising in the present model.

At orders higher than one the ST identities are intrinsically non-linear.
Nevertheless, by using the ST parameterization of the
quantum effective action it can be proven that in the model under investigation
there exists a special choice of normalization conditions
such that the relevant action-like part of the symmetric
quantum effective action obeys a linear identity also at higher
orders.
The existence of this special set of normalization conditions
is related to the geometrical structure of the non-invariant 
action-like terms.

By imposing these normalization conditions explicit formulas
for the action-like part of the symmetric vertex functional
are finally derived to all orders in the loop expansion.

We remark that the inclusion of composite operators like
the singlet axial-vector currents does not pose any additional problem
in the application of the ST parameterization technique.
Hence the latter also provides an efficient way to treat BRST-invariant 
composite operators in those theories where no invariant regularization
scheme is known. 

\medskip
The paper is organized as follows. In Sect.~\ref{sec:IR} 
the classical action of Yang-Mills theory with massive fermions 
in the presence of singlet axial-vector currents is considered
in the BPHZL scheme. The IR regulator for gauge fields and ghosts is
cast in a form useful for dealing with BRST symmetry
along the lines of \cite{Quadri:2003pq}.
In Sect.~\ref{sec:oneloop} the ST parameterization of the
quantum effective action is used in order to
derive the most general form of the action-like part of the
symmetric vertex functional. 
The fulfillment of the ST identities implies a set of
relations involving only superficially convergent Feynman amplitudes.
In Sect.~\ref{sec:WZ} we show that these relations 
actually hold true as a consequence of the Wess-Zumino
consistency condition.
In Sect.~\ref{sec:higher} the $n$-th order ST identities are analyzed
and a natural set of normalization conditions is obtained,
 leading to the linearization of the ST identity obeyed
by the action-like part of the $n$-th order symmetric vertex functional.
Finally conclusions are presented in Sect.~\ref{sec:concl}.

\section{The model}\label{sec:IR}

We consider a semi-simple gauge group $G$ and denote by $T^a$ the
hermitean generators of the associated Lie algebra 
$\mathfrak{g}$ in the fermionic matter representation. 
The generators $T^a$ obey the commutation relation
\begin{eqnarray}
[T^a,T^b]= i f^{abc} T^c 
\label{mod.1}
\end{eqnarray}
where $f^{abc}$ are the structure constants of $\mathfrak{g}$.

We wish to study the model described by the action:
\begin{eqnarray}
S & = & \int d^4x \, \Big ( -\frac{1}{4g^2} F^a_{\mu\nu} F^{\mu\nu a} 
          + i \bar \psi_i \slsh{D} \psi_i - M \bar \psi_i \psi_i 
\nonumber \\
& & ~~~~~~~~~~~~~~
+ \beta^V_\mu \bar \psi \gamma^\mu \psi + 
                    \beta^A_\mu \bar \psi \gamma^\mu \gamma^5 \psi \Big )
\, ,
\label{l1}
\end{eqnarray}
invariant under the following infinitesimal gauge transformations of 
parameters $\alpha^a$:
\begin{eqnarray}
&&  \delta A^a_\mu = (D_\mu \alpha)^a \equiv \partial_\mu \alpha^a + f^{abc} A^b_\mu \alpha^c \, , ~~~~~
\nonumber \\
&&  \delta \psi_i = i \alpha^a T^a \psi_i \, , ~~~~~
    \delta  \bar \psi_i =  -i \alpha^a \bar \psi_i T^a  \, .
\label{gauge.1}
\end{eqnarray}
In eq.(\ref{l1}) $g$ denotes the Yang-Mills coupling constant.
$F_{\mu\nu}^a$ stands for the field strength of the gauge fields $A_\mu^a$:
\begin{eqnarray}
F_{\mu\nu}^a = \partial_\mu A_\nu^a - \partial_\nu A_\mu^a + f^{abc} A_\mu^b
A_\nu^c \, .
\label{fs.1}
\end{eqnarray}
The covariant derivative on the fermionic fields $\psi_i$ is defined by
\begin{eqnarray}
D_\mu \psi_i = \partial_\mu \psi_i - i A_\mu^a T^a \psi_i \, .
\label{der_cov}
\end{eqnarray}
$\beta^V_\mu$  is an external source coupled to 
the singlet vector current $j^\mu_V = \bar \psi \gamma^\mu \psi$
while  the external source $\beta^A_\mu$ is coupled 
to the singlet axial current
$j^\mu_A = \bar \psi \gamma^\mu \gamma^5 \psi$.

The BRST transformations of $A_\mu^a$ and $\psi_i,\bar \psi_i$
are obtained in the usual way by replacing 
in the gauge transformations 
of eq.(\ref{gauge.1}) the gauge parameters $\alpha^a$
by the anticommuting 
ghost fields $\omega^a$. Nilpotency of the BRST differential $s$ 
fixes the BRST transformation of $\omega^a$. The action of $s$ on
$A_\mu^a, \bar \psi_i, \psi_i$ and $\omega^a$ is given 
in Appendix \ref{app:brst}. The external sources
$\beta^V_\mu$ and $\beta^A_\mu$ are BRST-invariant.
$S$ in eq.(\ref{l1}) is invariant under $s$.

In order to fix the gauge we choose a standard
Lorentz-covariant gauge-fixing condition
\begin{eqnarray}
F^a = \partial A^a \, , 
\label{l7}
\end{eqnarray}
which we implement in the BRST formalism 
by introducing the antighost fields $\bar \omega^a$
and the corresponding Nakanishi-Lautrup multiplier fields
$B^a$, with the following BRST transformations:
\begin{eqnarray}
s \bar \omega^a = B^a \, , ~~~~~ s B^a = 0 \, .
\label{l8}
\end{eqnarray}
Then the gauge-fixing term is
\begin{eqnarray}
S_{\textrm{g.f.}} & = & \int d^4x \, s \Big ( \bar \omega^a
\Big ( \alpha \frac{B^a}{2} - \partial A^a \Big ) \Big ) \nonumber \\
                  & = & \int d^4x \, \Big (
\alpha \frac{(B^a)^2}{2} - B^a \partial A^a + 
\bar \omega^a \partial^\mu ( D_\mu \omega)^a \Big ) \, .
\label{l9}
\end{eqnarray}
$\alpha$ is the gauge parameter.
Since the BRST transformations of $A_\mu^a, \psi_i,\bar \psi_i$
and $\omega^a$ are non-linear in the quantum fields, their
antifields \cite{Piguet:1995er} have to be introduced. We denote them
respectively by $A_\mu^{a*}, \bar Y_i, Y_i$ and $\omega^{a*}$.
The antifield-dependent part of the classical action is
obtained by coupling the antifields to the BRST variation
of the corresponding fields, as follows:
\begin{eqnarray}
S_{\textrm{a.f.}} & = & \int d^4x \, \Big (
A_\mu^{a*} (D^\mu \omega)^a
+ \omega^{a*} (-\frac{1}{2} f^{abc} \omega^b \omega^c ) \nonumber \\
& & ~~~~~~~~~
- i \bar \psi_i T^a \omega^a Y_i
+ i \bar Y_i \omega^a T^a \psi_i \Big ) \, .
\label{l10}
\end{eqnarray}
The full gauge-fixed BRST-invariant classical action is finally given by
\begin{eqnarray}
\G^{(0)} = S + S_{\textrm{g.f.}} + S_{\textrm{a.f.}}
\label{l11}
\end{eqnarray}

\subsection{IR regulator in the BPHZL scheme}

In the BPHZL regularization scheme
\cite{Bogolyubov:nc}-\cite{Lowenstein:1975rg}
massless propagators
are handled by the Lowenstein-Zimmermann prescription
\cite{Lowenstein:1975ps}-\cite{Lowenstein:1975rg}. 
An infrared regulator is introduced by assigning to all massless particles 
an intermediate mass
\begin{eqnarray}
m^2 = \mu^2 (s-1)^2
\label{ir.1}
\end{eqnarray}
The IR regulator $m^2$ depends on the auxiliary parameter $s$
ranging between $0$ and $1$ while $\mu$ is a fixed constant mass parameter.
The subtraction operator
$t_\gamma$ for a given divergent 1-PI graph or subgraph $\gamma$ involves
both a subtraction around $p=0,s=0$ and around $p=0,s=1$
\cite{Lowenstein:1975ps}-\cite{Lowenstein:1975rg}:
\begin{eqnarray}
(1 - t_\gamma) = (1 -t^{\rho(\gamma)-1}_{p,s-1})(1-t^{\delta(\gamma)}_{p,s}) 
\, ,
\label{ir.2}
\end{eqnarray}
where $\rho(\gamma)$ is the IR subtraction degree  and
$\delta(\gamma)$ the UV subtraction degree for $\gamma$
\cite{Lowenstein:1975ps}-\cite{Lowenstein:1975rg}.
We point out that both subtractions around $s=0$ 
and $s=1$ are needed in order to
guarantee the absence of IR singularities of the 1-PI Green functions
in the physical limit $s \rightarrow 1$ ($m \rightarrow 0$),
under the assumption that the IR power-counting criteria
of \cite{Lowenstein:1975ps}-\cite{Lowenstein:1975rg} are fulfilled.

The interplay between the ST identities and the BPHZL IR regulator
for massless gauge fields and ghosts \cite{Lowenstein:pd}
has been discussed from a cohomological point of view
in \cite{Quadri:2003pq}. The mass $m$ in eq.(\ref{ir.1}) is paired
within a BRST doublet with a constant anticommuting parameter $\bar \rho$
in such a way that
\begin{eqnarray}
&& s \bar \rho = m^2 \, , ~~~~ s m^2 = 0 \, .
\label{ir.3}
\end{eqnarray}
Notice that each factor $(s-1)$ (and hence $m$) in eq.(\ref{ir.1})
is assigned UV dimension one in the BPHZL 
subtraction scheme
\cite{Lowenstein:ug}
(unlike the fermionic mass $M$, which has UV dimension zero due to the fact
that no
subtraction in $M$ is performed).
The UV dimension of $\bar \rho$ is equal to one.

Then the following BRST-invariant mass term is added 
to the action $\G^{(0)}$ in eq.(\ref{l11}):
\begin{eqnarray}
&& \!\!\!\!\!\!\!\!\!\!\!\!\!\!\!\!\!\!\!\!\!\!\!\!\!\! s \int d^4x \, \Big (
\frac{1}{2}\bar \rho (A_\mu^a)^2 + \bar \rho \bar \omega^a \omega^a 
\Big ) 
= \int d^4x \, \Big ( \frac{1}{2} m^2 (A_\mu^a)^2 + m^2 \bar \omega^a
\omega^a  \nonumber \\
&& 
~~~~
- \bar \rho A_\mu^a \partial^\mu \omega^a - \bar \rho B^a \omega^a 
- \frac{1}{2} \bar \rho \bar \omega^a f^{abc} \omega^b \omega^c \Big ) \, .
\label{ir.4}
\end{eqnarray}
The cohomology of the original BRST differential is unaltered
by the inclusion of the BRST doublet $(\bar \rho,m^2)$.
Accordingly the mass term in eq.(\ref{ir.4}) is BRST-exact.
In the physical limit  $m \rightarrow 0, \bar \rho \rightarrow 0$
all the invariants of the BRST differential, extended according
to eq.(\ref{ir.3}), reduce to those of the original
BRST transformation.

The IR-regulated classical action $\G^{(0)}_m$ is reported
in Appendix \ref{app:brst}.
The ST identities for $\G^{(0)}_m$ become 
\begin{eqnarray}
{\cal S}(\G^{(0)}_m) & = & \int d^4x \, \Big (
\frac{\delta \G^{(0)}_m}{\delta A_\mu^{a*}} \frac{\delta \G^{(0)}_m}{\delta A^{a\mu}}
+
\frac{\delta \G^{(0)}_m}{\delta \omega^{a*}} \frac{\delta \G^{(0)}_m}{\delta \omega^a}
+
B^a \frac{\delta \G^{(0)}_m}{\delta \bar \omega^a}
\nonumber \\
& & ~~~~~~~~~
- \frac{\delta \G^{(0)}_m}{\delta Y_i} 
  \frac{\delta \G^{(0)}_m}{\delta \bar \psi_i} 
+ \frac{\delta \G^{(0)}_m}{\delta \bar Y_i} 
  \frac{\delta \G^{(0)}_m}{\delta \psi_i} 
\Big ) + m^2 \frac{\partial \G^{(0)}_m}{\partial \bar \rho} = 0 \, .
\label{ir.5}
\end{eqnarray}
The last term in the above equation parameterizes the soft-breaking of 
the original ST identities at the intermediate regularized level. 
Explicit violations of Physical Unitarity 
for $m \neq 0$ can be studied e.g. by using the technique of \cite{Picariello:2001ri}.

The classical linearized ST operator \cite{Piguet:1995er} is 
\begin{eqnarray}
{\cal S}_0 & = & \int d^4x \, 
\Big ( (D_\mu \omega)^a \frac{\delta}{\delta A_\mu^a} 
     -\frac{1}{2} f^{abc} \omega^b \omega^c \frac{\delta}{\delta \omega^a}
     + B^a \frac{\delta}{\delta \bar \omega^a} \nonumber \\
       &   & ~~~~~~~~~ + i \omega^a T^a \psi_i \frac{\delta}{\delta \psi_i}
                    + i \bar \psi_i T^a \omega^a \frac{\delta}{\delta \bar \psi_i}
       \nonumber \\
       &   & ~~~~~~~~~ + \frac{\delta \G^{(0)}_m}{\delta A^{a\mu}} \frac{\delta}{\delta A_\mu^{a*}}
                    + \frac{\delta \G^{(0)}_m}{\delta \omega^a} \frac{\delta}{\delta \omega^{a*}}
       \nonumber \\
       &   & ~~~~~~~~~ + \frac{\delta \G^{(0)}_m}{\delta \psi_i} \frac{\delta}{\delta \bar Y_i}
                    - \frac{\delta \G^{(0)}_m}{\delta \bar \psi_i} \frac{\delta}{\delta Y_i}
\Big ) +  m^2 \frac{\partial}{\partial \bar \rho} \, .
\label{l13}
\end{eqnarray}
${\cal S}_0$ is nilpotent as a consequence of eq.(\ref{ir.5}).

\subsection{Additional symmetries of the model}

The dependence of the classical action $\G^{(0)}_m$  
on the Nakanishi-Lautrup multiplier field $B^a$ and on 
the antighost $\bar \omega^a$
is controlled by the following set of identities:
\begin{itemize}
\item the $B$-equation:
\begin{eqnarray}
\frac{\delta \G^{(0)}_m}{\delta B^a} = \alpha B_a - \partial A_a - 
\bar \rho  \omega_a \, ;
\label{l14}
\end{eqnarray}
\item the antighost equation:
\begin{eqnarray}
\frac{\delta \G^{(0)}_m}{\delta \bar \omega_a} = 
\partial_\mu \frac{\delta \G^{(0)}_m}{\delta A_\mu^{a*}} 
- \bar \rho \frac{\delta \G^{(0)}_m}{\delta \omega^{a*}} 
+ m^2 \omega^a
\, .
\label{l15}
\end{eqnarray}
\end{itemize}
Both relations can be translated at the quantum level. Therefore 
the symmetric quantum vertex functional $\GG_m$ satisfies
\begin{eqnarray}
&& \frac{\delta \GG_m}{\delta B^a} = \alpha B_a - \partial A_a - 
\bar \rho  \omega_a \, , 
\label{l14_quant} \\
&& \frac{\delta \GG_m}{\delta \bar \omega_a} = 
\partial_\mu \frac{\delta \GG_m}{\delta A_\mu^{a*}} 
- \bar \rho \frac{\delta \GG_m}{\delta \omega^{a*}} 
+ m^2 \omega^a
\, .
\label{l15_quant}
\end{eqnarray}
We expand $\GG_m$ in powers of $\hbar$ as follows:
\begin{eqnarray}
\GG_m = \sum_{n=0}^\infty \GG_m^{(j)}
\label{exp.1}
\end{eqnarray}
where $\GG_m^{(j)}$ denotes the coefficient of $\GG_m$ of order $j$
in the loop expansion. $\GG_m^{(0)}$ coincides with $\G^{(0)}_m$.
By projecting eq.(\ref{l14_quant}) at order $j \geq 1$ it follows 
that $\GG^{(j)}_m$ is $B^a$-independent. Moreover, by projection of 
eq.(\ref{l15_quant}) at order $j \geq 1$ $\GG^{(j)}_m$ is seen to depend
on $\bar \omega_a$ only via the combinations
\begin{eqnarray}
A^{a*'}_{\mu} = A^{a*}_\mu - \partial_\mu \bar \omega^a \, , ~~~~
\omega^{a*'} = \omega^{a*} + \bar \rho \bar \omega^a \, .
\label{exp.2}
\end{eqnarray}
Their variations under ${\cal S}_0$ are
\begin{eqnarray}
&& {\cal S}_0 (A^{a*'}_{\mu}) = \left . \frac{\delta \G^{(0)}_m}{\delta A^a_\mu}
\right |_{B^a=0} \, , \nonumber \\
&& {\cal S}_0 (\omega^{a*'}) = \left . \frac{\delta \G^{(0)}_m}{\delta \omega^a}
\right |_{B^a=0} + m^2 \bar \omega^a 
= \left . \frac{\delta \G^{(0)}_m}{\delta \omega^a}
\right |_{B^a=0,m^2=0}
\, .
\label{exp.3}
\end{eqnarray}

\section{One-loop ST identities}\label{sec:oneloop}
At one loop level the ST identities are
\begin{eqnarray}
{\cal S}_0 (\GG^{(1)}_m) = 0 \, .
\label{exp.4}
\end{eqnarray}
It is useful to isolate in ${\cal S}_0$ the term acting on
$\bar \rho$. Thus we define the differential $\delta$ according to
\begin{eqnarray}
{\cal S}_0 = \delta + m^2 \frac{\partial}{\partial \bar \rho} \, .
\label{exp.5}
\end{eqnarray}
Since the counting operator ${\cal N} = \bar \rho \frac{\partial}{\partial
  \bar \rho} + m^2 \frac{\partial}{\partial m^2}$ does not commute
with ${\cal S}_0$ $(\bar \rho, m^2)$ form a set of BRST coupled doublets
\cite{Quadri:2002nh}. As such they do not contribute to the
cohomology of ${\cal S}_0$. In the present model it is possible
to remove the dependence of $\delta$ on $(\bar \rho, m^2)$
by a suitable change of coordinates 
\cite{Brandt:1996mh}-\cite{Brandt:2001tg},\cite{Barnich:2000zw}.
The appropriate redefinition is given by
\begin{eqnarray}
\hat A^{a*}_\mu = A^{a*'}_\mu - \bar \rho A^a_\mu \, , ~~~~
\hat \omega^{a*} = \omega^{a*'} \, .
\label{exp.6}
\end{eqnarray}
One finds
\begin{eqnarray}
&& {\cal S}_0 (\hat A^{a*}_\mu) = \left . \frac{\delta \G^{(0)}_m}{\delta A_\mu^a}
\right |_{B^a=0,m^2=0} \, , \nonumber \\
&& {\cal S}_0 (\hat \omega^{a*}) = \left . \frac{\delta \G^{(0)}_m}{\delta \omega^a}
\right |_{B^a=0,m^2=0} \, ,
\label{exp.7}
\end{eqnarray}
where the R.H.S. in the above equation is expressed
in terms of $\hat A^{a*}_\mu, \hat \omega^{a*}$. We notice that 
the ${\cal S}_0$-variations of $\hat A^{a*}_\mu,\hat \omega^{a*}$
are in the new variables $\bar \rho$- and $m^2$-independent.
Therefore in the new variables $\delta$ commutes with both 
${\cal N}_{\bar \rho} = \bar \rho \frac{\partial}{\partial \bar\rho}$
and ${\cal N}_{m^2} = m^2 \frac{\partial}{\partial m^2}$.
From the nilpotency of ${\cal S}_0$ and the fact that
$\{ m^2 \frac{\partial}{\partial \bar \rho}, \delta \}=0$
it follows that $\delta$ is also nilpotent.

The existence of the set of coordinates in eq.(\ref{exp.6}) has
to be traced back to the fact that $(\bar \rho, m^2)$
enter in $\G^{(0)}_m$ in a cohomologically trivial way.
%
%
%\medskip
From now on we will work with the hatted variables in eq.(\ref{exp.6}).

\subsection{Construction of the one-loop solution}

The presence of the IR regulator $m$ in eq.(\ref{ir.1})
allows to perform the expansion of the 1-PI Green functions
around zero momentum. 
We can then define the projector $t^4$ on the action-like part
of $\GG^{(1)}_m$ as follows. Let $\Phi_I$, $I=1,\dots,N$ stand
for the fields, antifields and external sources of the model and
let $d(\Phi_I)$ be the UV dimension of $\Phi_I$.
For any $n$-tuple $\underline{j} = \{ j_1,\dots, j_n \}$
we set $d(\underline{j}) = \sum_{k=1}^n d(\Phi_{j_k})$ 
and $|\underline{j}| = n$, the length of $\underline{j}$.
Then
\begin{eqnarray}
&& 
\!\!\!\!\!\!\!\!\!\!\!\!\!\!
t^4 \GG^{(1)}_m = 
\sum_{n=1}^\infty 
\sum_{\scriptsize \matrix{|\underline{j}|=n, \cr d(\underline{j}) \leq 4}} 
\int d^4p_1 \dots d^4p_n \, \delta^{(4)}(p_1 + \dots + p_n) \nonumber \\
&& ~~~~~~ \times 
t^{4 - d(\underline{j}) }
\left . \frac{\delta^{(n)} \GG^{(1)}_m}{\delta\Phi_{j_n}(p_n) \dots
  \delta\Phi_{j_1}(p_1)} \right |_{\Phi=0} 
\Phi_{j_n}(p_n) \dots \Phi_{j_1}(p_1)
\, 
\label{st.brkg.1}
\end{eqnarray}
where $t^{4 - d(\underline{j})}$ is the Taylor operator around $p_i=0$ 
up to dimension $4 - d(\underline{j})$ in the 
independent momenta $p_i$, $i=1,\dots,n-1$.

In the configuration space $t^4 \GG^{(1)}_m$ is the projection
of $\GG^{(1)}_m$ on the sector spanned by
all Lorentz-invariant monomials of dimension
$\leq 4$ in the fields, the antifields, the external sources
and their derivatives, compatible with the unbroken
symmetries of the regularized action.

By the Quantum Action Principle \cite{Breitenlohner:1977hr}-\cite{Lam:1973qk}
the one-loop BPHZL regularized, generally non-symmetric vertex functional
$\G^{(1)}_m$ may spoil the ST identities only by a local functional
of dimension $\leq 5$ in the fields, the antifields, the external sources and
their derivatives.
Therefore in order to study the one-loop ST identites
\begin{eqnarray}
{\cal S}_0 (\GG^{(1)}_m) = 0
\label{eq.1}
\end{eqnarray}
we can replace $\GG^{(1)}_m$ with its effective part (i.e. the Taylor
expansion of any amplitude in the independent external momenta
around zero). This procedure associates to $\GG^{(1)}_m$ a local
formal power series given by an infinite sum of local Lorentz-invariant
functionals. Without possibility of confusion we can 
denote this series by $\GG^{(1)}_m$ itself.

Moreover, in the absence of anomalies this breaking can be compensated
by adding to $\G^{(1)}_m$ a local functional
$\Upsilon^{(1)}$ in the fields, the antifields, the external sources and
their derivatives of dimension $\leq 4$, in such a way that
\begin{eqnarray}
\GG^{(1)}_m = \G^{(1)}_m + \Upsilon^{(1)}, ~~~~ 
t^4 \Upsilon^{(1)} = \Upsilon^{(1)} \, .
\label{corr.1}
\end{eqnarray}

By the above equation it follows that 
\begin{eqnarray}
(1- t^4) \GG^{(1)}_m = (1 - t^4)  \G^{(1)}_m \, .
\label{corr.2}
\end{eqnarray}
Therefore the regularization procedure yields the correct
symmetric Green functions in the whole sector of dimension $>4$.
The only unknown part is the action-like part $t^4 \GG^{(1)}_m$.
Once $\GG^{(1)}_m$ is known $\Upsilon^{(1)}$ can be
read off by comparison with eq.(\ref{corr.1}).

Eq.(\ref{eq.1}) defines a cohomological problem in the space of local
formal power series spanned by $A_\mu^a, \bar \psi, \psi, \omega^a,
\hat \omega^{a*}, \hat A^{a*}_\mu$, $Y_i, \bar Y_i, \bar \rho, m^2$, $\beta_\mu^V, \beta_\mu^A$
and their derivatives (without power-counting restrictions). Since
we are interested only in $t^4 \GG^{(1)}_m$ we rewrite eq.(\ref{eq.1}) as
\begin{eqnarray}
{\cal S}_0 (t^4 \GG^{(1)}_m) = -{\cal S}_0 ((1-t^4)\GG^{(1)}_m) \, .
\label{eq.1.bis}
\end{eqnarray}

It turns out \cite{Quadri:2003ui} that an effective way
to find the most general solution of eq.(\ref{eq.1.bis})
is to parameterize the L.H.S. in terms of ST invariants 
(functionals belonging to the kernel of ${\cal S}_0$).
On the other hand, the R.H.S. is expanded on a basis of monomials 
in the fields, the external sources and their derivatives. 
The coefficients of these monomials are given in terms of known
superficially convergent vacuum amplitudes.
Then eq.(\ref{eq.1.bis}) is used to obtain directly 
the functional $t^4 \GG^{(1)}_m$.

In order to apply this technique 
it is useful to decompose ${\cal S}_0$ according to the degree induced
by the UV dimension.
By inspection it is seen that $\delta$ 
splits into two components of degree zero and one respectively:
\begin{eqnarray}
\delta = \delta_0 + \delta_1 \, ,
\label{decomp.1}
\end{eqnarray}
while $m^2 \frac{\partial}{\partial \bar \rho}$ has degree one.
$\delta_0$ keeps the dimension constant, while $\delta_1$ and 
$m^2 \frac{\partial}{\partial \bar \rho}$ increase
it by one.
The action of $\delta_0$ and $\delta_1$ on the variables
of the model is given in Appendix ~\ref{app:decomp}.

\subsection{ST parameterization of $t^4 \GG^{(1)}_m$}

By eq.(\ref{eq.1}) $\GG^{(1)}_m$ belongs to the kernel of
${\cal S}_0$. The most general form of an arbitrary
${\cal S}_0$-invariant is known \cite{Barnich:1994ve}-\cite{Barnich:1994mt} 
and is given by
\begin{eqnarray}
\GG^{(1)}_m = \sum_j \lambda_j^{(1)} \Lambda_j + \sum_k \rho^{(1)}_k {\cal S}_0(R_k)
\, ,
\label{sol.1}
\end{eqnarray}
where $\Lambda_j$ are ${\cal S}_0$-invariants that are not
${\cal S}_0$-exact (i.e. they cannot be expressed
as ${\cal S}_0$-variations of a functional $\Theta_j$) and
${\cal S}_0(R_k)$ are the trivial solutions (they are in the kernel
of ${\cal S}_0$ since ${\cal S}_0$ is nilpotent).

For a semi-simple gauge group $G$
$\Lambda_j$ are gauge-invariant polynomials built from the
field strength $F_{\mu\nu}^a$ and its covariant derivatives,
the matter fields and their covariant derivatives and the sources
$\beta_V^\mu, \beta_A^\mu$, while all the dependence on the antifields
and on $\bar \rho$ is confined to the trivial invariants.

If one wishes to parameterize the full $\GG^{(1)}_m$ the whole
infinite tower of functionals $\Lambda_j$ and ${\cal S}_0(R_k)$
is needed. However, since we are interested only in 
$t^4 \GG^{(1)}_m$, a finite number of them has to be considered.
All functionals $\Lambda_j$ and ${\cal S}_0(R_k)$ such that
$t^4 \Lambda_j = t^4 {\cal S}_0(R_k)=0$ can be dropped.

On the other hand the invariants contributing to the action-like sector
split into two distinct subsets.
The first one is generated by those invariants $X$ such that
$t^4 X \neq 0$, $(1-t^4)X=0$. The coefficients of these invariants 
are not constrained by the ST identities and have to be 
fixed by the choice of a complete set of normalization conditions.
In addition there may also appear invariants
such that $t^4 X \neq 0$, $(1-t^4)X \neq 0$.
Their coefficients are not free but fixed by superficially convergent
Feynman amplitudes.
They typically appear in the presence of cohomologically
non-trivial mass parameters. In this model 
they must be of the trivial form (i.e. they are
${\cal S}_0$-exact) and fulfill
$t^4 {\cal S}_0(R_k) \neq 0 \, , ~ (1-t^4) {\cal S}_0(R_k) \neq 0$.
They contain at least
one $Y_i, \bar Y_i$ because the only component 
of ${\cal S}_0$ which does not increase the UV dimension is 
$\delta_0$ and the latter acts non-trivially only on the fermionic
antifields $\bar Y_i, Y_i$.

The presence in the model of invariants of this kind is the reason
why non-trivial non-symmetric terms, depending on
superficially convergent amplitudes, have to be included 
in the action-like sector of $\GG^{(1)}_m$ in order
to fulfill the ST identities.

\medskip
Our next step is to find the complete set of invariants needed
to parameterize $t^4 \GG^{(1)}_m$ (i.e. such that their
action-like part is non-vanishing). 

We first list those invariants $X$ such that 
their non-action-like part $(1-t^4)X$ vanishes.
They can be classified according to the following groups:
\begin{itemize}
\item cohomologically non-trivial invariants independent
of $\beta_V^\mu,\beta_A^\mu$:
\begin{eqnarray}
&& \Lambda_1 = \int d^4x \, F_{\mu\nu}^a F^{\mu\nu \, a} \, , \nonumber \\
&& \Lambda_2 = \int d^4x \, \bar \psi_i \psi_i \, .
\label{st1}
\end{eqnarray}
We notice that the covariant kinetic term for the fermions,
\begin{eqnarray}
\int d^4x \, i \bar \psi_i \slsh{D} \psi_i \, , 
\label{cov.kin}
\end{eqnarray}
is an independent BRST-invariant but not an independent
${\cal S}_0$-invariant (see eq.(\ref{av.ol2}) and comments
thereafter).
\item cohomologically non-trivial invariants depending on 
$\beta_V^\mu,\beta_A^\mu$:
\begin{eqnarray}
&& \Lambda_V = \int d^4x \, \beta_\mu^V \bar \psi \gamma^\mu \psi \, ,
\nonumber \\
&& \Lambda_A = \int d^4x \, \beta_\mu^A \bar \psi \gamma^\mu \gamma^5 \psi \, ,
\label{betava}
\end{eqnarray}
plus sixteen invariants only dependent on $\beta_V^\mu,\beta_A^\mu$
listed in Appendix~\ref{app:beta_inv}.
\item cohomologically trivial invariants depending on $\bar\rho$

There is just one invariant of this kind:
\begin{eqnarray}
R_{\bar \rho} = {\cal S}_0(\int d^4x \, \frac{\bar \rho}{2} (A_\mu^a)^2) =
\frac{m^2}{2} \int d^4x \, (A_\mu^a)^2 - 
\int d^4x \, \bar \rho A_\mu^a \partial^\mu \omega^a \, .
\label{barrho.inv}
\end{eqnarray}
\item cohomologically trivial invariants independent of $\bar\rho$:
\begin{eqnarray}
\!\!\!\!\!\!\!
R_1 & = & \delta ( \int d^4x \, \hat A_\mu^{a*} A^\mu_a ) \, , \nonumber \\
\!\!\!\!\!\!\!
R_2 & = & \delta ( \int d^4x \, \hat \omega^*_a \omega_a) \, , \nonumber \\
\!\!\!\!\!\!\!
R_3 & = & \frac{1}{2} \delta \int d^4x \, 
                 \Big ( \bar \psi_i Y_i - \bar Y_i \psi_i \Big ) \nonumber \\
           & = & \int d^4x \, 
                 \Big ( i \bar \psi_i \slsh{D} \psi_i - M \bar \psi_i \psi_i
                 + \beta_\mu^V \bar \psi_i \gamma^\mu \psi_i
                 + \beta_\mu^A \bar \psi_i \gamma^\mu \gamma^5 \psi_i 
                 \Big ) \, . 
\label{av.ol2}
\end{eqnarray}
We notice that by the above equation the BRST-invariant
$$\int d^4x \, i \bar \psi_i \slsh{D} \psi_i$$ is in the same
cohomology class of the sum 
$$\int d^4x \, \Big ( - M \bar \psi_i \psi_i
                 + \beta_\mu^V \bar \psi_i \gamma^\mu \psi_i
                 + \beta_\mu^A \bar \psi_i \gamma^\mu \gamma^5 \psi_i 
               \Big ) \, .$$
Therefore either $R_3$ or $\int d^4x \,  i \bar \psi_i \slsh{D} \psi_i$
has to be included in the parameterization of $t^4 \GG^{(1)}_m$. 
In what follows
we choose to use $\int d^4x \,  i \bar \psi_i \slsh{D} \psi_i$
in order to achieve a direct control of the normalization condition
for the residue of the fermionic field propagator.
\end{itemize}
The coefficients of the invariants in eqs.(\ref{st1}), (\ref{betava}),
(\ref{barrho.inv}) and (\ref{av.ol2}) are free and have to be fixed
by providing a set of normalization conditions.

\medskip
Let us consider now those invariants $X$ such that 
$t^4 X \neq 0$ and $(1-t^4)X \neq 0$. They have to be ${\cal S}_0$-trivial
and contain at least one $\bar Y_i, Y_i$. By power-counting
they cannot depend on $\bar \rho$.
The relevant invariants are
\begin{eqnarray}
R_4 & = & \frac{1}{2} \delta ( \int d^4x \, \bar Y_i \slsh{A}^a T^a \psi_i 
- \int d^4x \, \bar \psi_i T^a \slsh{A}^a Y_i ) \, , \nonumber \\
R_5 & = & \frac{1}{2} \delta ( \int d^4x \, i \bar Y_i \slsh{\partial} \psi_i 
- \int d^4x \, ( - i \slsh{\partial} \bar \psi_i Y_i ) ) \, , \nonumber \\
R_6 & = & \delta  \int d^4x \, i \bar Y_i T^a \omega^a Y_i \, , \nonumber \\
R_V & = & \delta \int d^4x \, 
\Big ( \beta_\mu^V \bar Y_i \gamma^\mu \psi_i 
      -\beta_\mu^V \bar \psi_i \gamma^\mu Y_i \Big ) \, , \nonumber \\  
R_A & = & \delta \int d^4x \,
\Big ( \beta_\mu^A \bar Y_i \gamma^\mu \gamma^5 \psi_i 
      -\beta_\mu^A \bar \psi_i \gamma^\mu \gamma^5 Y_i \Big ) \, .
\label{inv_nal}
\end{eqnarray}
Unlike the coefficients of the invariants in eqs.
(\ref{st1}),(\ref{betava})-(\ref{av.ol2})
the coefficients of the invariants in eq.(\ref{inv_nal})
are fixed by superficially convergent Feynman amplitudes.

\medskip
By making use of these ST invariants the functional $t^4 \GG^{(1)}_m$ solving
eq.(\ref{eq.1.bis}) can be written as
\begin{eqnarray}
t^4 \GG^{(1)} & = & \lambda^{(1)}_1 \int d^4x \, F_{\mu\nu}^a F^{\mu\nu a}
                   +\lambda^{(1)}_2 \int d^4x \, \bar \psi_i \psi_i
                   +\rho_3^{(1)} \int d^4x \, i \bar \psi_i \slsh{D} \psi_i
\nonumber \\
&   & + \sum_{i=1}^{16} \lambda^{(1)}_{\beta,i} \Lambda_{\beta,i}(x)
      + \lambda_V^{(1)}  \int d^4x \, \beta^V_\mu \bar \psi
                   \gamma^\mu \psi 
      + \lambda_A^{(1)}  \int d^4x \, \beta^A_\mu \bar \psi
                   \gamma^\mu \gamma^5 \psi 
\nonumber \\
&   & +\rho^{(1)}_{\bar \rho} \Big ( \frac{m^2 }{2} \int d^4x \, (A_\mu^a)^2 - 
\int d^4x \, \bar \rho A_\mu^a \partial^\mu \omega^a \Big ) \nonumber \\
&   & +\rho^{(1)}_1 \delta \int d^4x \, (\hat A^{a*}_\mu A^{a\mu})
      +\rho^{(1)}_2 \delta \int d^4x \, (\hat \omega^{a*} \omega^a )
\nonumber \\
&   & + \frac{1}{2} \rho_4^{(1)} t^4 \delta \int d^4x \, 
(\bar Y_i \slsh{A}^a T^a \psi_i - \bar \psi_i T^a \slsh{A}^a Y_i) 
\nonumber \\
&   & + \frac{1}{2} \rho_5^{(1)} t^4 \delta \int d^4x \,
(i \bar Y_i \slsh{\partial} \psi_i + i \slsh{\partial} \bar \psi_i Y_i)
\nonumber \\
&   & + \rho_6^{(1)} t^4 \delta \int d^4x \, i \bar Y_i T^a \omega^a Y_i 
\nonumber \\
&   & + \frac{\rho^{(1)}_V}{2} t^4 \delta \int d^4x \, \Big (
\beta_\mu^V \bar Y_i \gamma^\mu \psi_i - 
\beta_\mu^V \bar \psi_i \gamma^\mu Y_i \Big ) \nonumber \\
&   & + \frac{\rho^{(1)}_A}{2} t^4 \delta \int d^4x \, \Big (
\beta_\mu^A \bar Y_i \gamma^\mu \gamma^5 \psi_i - 
\beta_\mu^A \bar \psi_i \gamma^\mu \gamma^5 Y_i \Big ) \nonumber \\
& = &  \lambda^{(1)}_1 \int d^4x \, F_{\mu\nu}^a F^{\mu\nu a}
                   +\lambda^{(1)}_2 \int d^4x \, \bar \psi_i \psi_i
                   +\rho_3^{(1)} \int d^4x \, i \bar \psi_i \slsh{D} \psi_i
+ \sum_{i=1}^{16} \lambda^{(1)}_{\beta,i} \Lambda_{\beta,i}(x)
\nonumber \\
& & 
      + ( \lambda_V^{(1)} + M \rho_V^{(1)}) \int d^4x \, \beta^V_\mu \bar \psi
                   \gamma^\mu \psi 
      + ( \lambda_A^{(1)}  + M \rho_A^{(1)} ) \int d^4x \, \beta^A_\mu \bar \psi
                   \gamma^\mu \gamma^5 \psi 
\nonumber \\
&   & +\rho^{(1)}_{\bar \rho} \Big ( \frac{m^2 }{2} \int d^4x \, (A_\mu^a)^2 - 
\int d^4x \, \bar \rho A_\mu^a \partial^\mu \omega^a \Big ) \nonumber \\
&   & +\rho^{(1)}_1 \delta \int d^4x \, (\hat A^{a*}_\mu A^{a\mu})
      +\rho^{(1)}_2 \delta \int d^4x \, (\hat \omega^{a*} \omega^a )
\nonumber \\
&   & + \rho_4^{(1)} M \int d^4x \, \bar \psi_i \slsh{A}^a T^a \psi_i
      + \rho_5^{(1)} M \int d^4x \, i \bar \psi_i \slsh{\partial} \psi_i
\nonumber \\
&   & + \rho^{(1)}_6 \int d^4x \,
( i M \bar \psi_i T^a \omega^a Y_i - i M \bar Y_i T^a \omega^a \psi_i)
\, .
\label{st9.av.1}
\end{eqnarray}
The dependence on $\rho^{(1)}_V, \rho^{(1)}_A$ can always be removed
by performing the shifts
\begin{eqnarray}
\lambda^{(1)}_V \rightarrow \lambda^{(1)}_V + M \rho_V^{(1)} \, , ~~~~
\lambda^{(1)}_A \rightarrow \lambda^{(1)}_A + M \rho_A^{(1)} \, ,
\label{shift.1}
\end{eqnarray}
which we assume from now on.

The free parameters of the solution are therefore
$\lambda^{(1)}_1,\lambda^{(1)}_2,\rho_3^{(1)}$,
$\lambda^{(1)}_{\beta,i}$,$\lambda_V^{(1)},\lambda_A^{(1)}$ and
$\rho_{\bar \rho}^{(1)},\rho^{(1)}_1, \rho^{(1)}_2$. On the contrary
$\rho_4^{(1)},  \rho_5^{(1)}$ and  $\rho^{(1)}_6$ are
fixed c-numbers (depending on superficially convergent Feynman amplitudes).

\subsection{Completing the construction}

In order to compute the fixed ST parameters 
$\rho_4^{(1)},  \rho_5^{(1)}$ and  $\rho^{(1)}_6$ entering into
eq.(\ref{st9.av.1}) we use eq.(\ref{eq.1.bis}).
We first compute the L.H.S. by applying ${\cal S}_0$ to 
eq.(\ref{st9.av.1}).
This yields
\begin{eqnarray}
{\cal S}_0 (t^4 \GG^{(1)}) & = & M ( (\rho_5^{(1)} - \rho_4^{(1)}) - \rho_6^{(1)})
 \int d^4x \,
\bar \psi_i \slsh{\partial} \omega^a T^a \psi_i  \nonumber \\
&  & - M \rho_6^{(1)} \int d^4x \, \bar \psi_i T^a f^{abc} \slsh{A}^b
 \omega^c \psi_i  \nonumber \\
&  & -  \frac{i M}{2} \rho_6^{(1)} \int d^4x \, ( 
\bar \psi_i T^a f^{abc} \omega^b \omega^c Y_i 
+ \bar Y_i T^a f^{abc} \omega^b \omega^c \psi_i ) \, .
\nonumber \\
\label{st16}
\end{eqnarray}
On the other hand we expand $(1-t^4) \GG^{(1)}_m$ in a sum of Lorentz-invariant
monomials of dimension $\geq 5$ in the fields, 
the antifields, the external sources
and their derivatives. We only need to consider those monomials
${\cal N}_j$ whose ${\cal S}_0$-variation contains the monomials 
appearing in eq.(\ref{st16}). 
The relevant six ${\cal N}_j$'s, whose coefficients we
denote by $\gamma^{(1)}_{{\cal N}_j(x)}$, 
are listed in Appendix~\ref{app:brkg_list}. 
By computing the ${\cal S}_0$-variation of $(1-t^4) \GG^{(1)}_m \approx
\sum_{j=1}^6 \gamma^{(1)}_{{\cal N}_j(x)} {\cal N}_j$ we finally get
\begin{eqnarray}
{\cal S}_0(\sum_{j=1}^6 \gamma^{(1)}_{{\cal N}_j(x)} {\cal N}_j)
& = &
i M (\gamma^{(1)}_{{\cal N}_1(x)} + \gamma^{(1)}_{{\cal N}_2(x)}
     - \gamma^{(1)}_{{\cal N}_3(x)} ) \int d^4x \,
\bar \psi_i \slsh{\partial} \omega^a T^a \psi_i \nonumber \\
&   & + M (\gamma^{(1)}_{{\cal N}_4(x)} + \gamma^{(1)}_{{\cal N}_5(x)}) 
\int d^4x \, \bar \psi_i T^a f^{abc} \slsh{A}^b \omega^c \psi_i
\nonumber \\
&   & + M \gamma^{(1)}_{{\cal N}_6(x)} 
\int d^4x \, i \bar\psi_i T^a f^{abc} \omega^b \omega^c Y_i
\nonumber \\
&   & + M \gamma^{(1)}_{{\cal N}_6(x)} 
\int d^4x \, i \bar Y_i T^a f^{abc} \omega^b \omega^c \psi_i \, .
\label{conv.brkg}
\end{eqnarray}
By using eq.(\ref{st16}), eq.(\ref{conv.brkg}) and eq.(\ref{eq.1.bis}) 
we obtain the following set of linear relations
\begin{eqnarray}
&& \rho_6^{(1)} = 2 \gamma^{(1)}_{{\cal N}_6(x)} = 
\gamma^{(1)}_{{\cal N}_4(x)} + \gamma^{(1)}_{{\cal N}_5(x)} \, , \label{st23.1}
\\
&& (\rho_5^{(1)} - \rho_4^{(1)}) - \rho_6^{(1)} = - i 
(\gamma^{(1)}_{{\cal N}_1(x)} + \gamma^{(1)}_{{\cal N}_2(x)}
- \gamma^{(1)}_{{\cal N}_3(x)} ) \, .
\label{st23}
\end{eqnarray}
Several comments are in order here. First we observe that
the set of conditions in eqs.(\ref{st23.1})-(\ref{st23}) allows to determine
$\rho_6^{(1)}$ and the difference $\rho_5^{(1)} - \rho_4^{(1)}$:
\begin{eqnarray}
\!\!\!\!\!\!\!\!\!\!
\rho_6^{(1)} = 2 \gamma^{(1)}_{{\cal N}_6(x)} \, , ~~~~
\rho_5^{(1)} - \rho_4^{(1)} =  2 \gamma^{(1)}_{{\cal N}_6(x)}
- i 
(\gamma^{(1)}_{{\cal N}_1(x)} + \gamma^{(1)}_{{\cal N}_2(x)}
- \gamma^{(1)}_{{\cal N}_3(x)} ) \, .
\label{st.comm.1}
\end{eqnarray}
By exploiting the freedom in the choice of $\rho^{(1)}_3$ we
redefine 
\begin{eqnarray}
\rho_3^{(1)} \rightarrow \rho_3^{(1)} - M \rho_5^{(1)} \, .
\label{st24}
\end{eqnarray}
Then $t^4 \GG_m^{(1)}$ in eq.(\ref{st9.av.1}) is seen
to depend on $\rho_4^{(1)}, \rho_5^{(1)}$ only via the monomial
\begin{eqnarray}
(\rho_3^{(1)}  - M (\rho_5^{(1)} -  \rho_4^{(1)}) ) 
\int d^4x \, \bar \psi_i \slsh{A}^a T^a \psi_i
\label{mon.1}
\end{eqnarray}
whose coefficient is controlled by the difference
$\rho_5^{(1)} - \rho_4^{(1)}$.
By using eqs.(\ref{st.comm.1}) we obtain finally the most general
solution for the action-like part of the symmetric vertex
functional $t^4 \GG_m^{(1)}$:
\begin{eqnarray}
t^4 \GG^{(1)}_m & = &
\lambda^{(1)}_1 \int d^4x \, F_{\mu\nu}^a F^{\mu\nu a}
                   +\lambda^{(1)}_2 \int d^4x \, \bar \psi_i \psi_i
\nonumber \\
&   & + \sum_{i=1}^{16} \lambda^{(1)}_{\beta,i} \Lambda_{\beta,i}(x)
      + \lambda_V^{(1)}  \int d^4x \, \beta^V_\mu \bar \psi
                   \gamma^\mu \psi 
      + \lambda_A^{(1)}  \int d^4x \, \beta^A_\mu \bar \psi
                   \gamma^\mu \gamma^5 \psi 
\nonumber \\
&   & +\rho^{(1)}_{\bar \rho} \Big ( \frac{m^2 }{2} \int d^4x \, (A_\mu^a)^2 - 
\int d^4x \, \bar \rho A_\mu^a \partial^\mu \omega^a \Big ) \nonumber \\
&   & +\rho^{(1)}_1 \delta \int d^4x \, (\hat A^{a*}_\mu A^{a\mu})
      +\rho^{(1)}_2 \delta \int d^4x \, (\hat \omega^{a*} \omega^a )
\nonumber \\
&   &  +\rho^{(1)}_3 \int d^4x \, i \bar \psi_i \slsh{\partial} \psi_i 
\nonumber \\
&  & + (\rho_3^{(1)}  
+ i M
(\gamma^{(1)}_{{\cal N}_1(x)} + \gamma^{(1)}_{{\cal N}_2(x)} - \gamma^{(1)}_{{\cal N}_3(x)} )
-2 M \gamma^{(1)}_{{\cal N}_6(x)} )
 \int d^4x \, \bar \psi_i \slsh{A}^a T^a \psi_i 
 \nonumber \\
&  & + 2 \gamma^{(1)}_{{\cal N}_6} \int d^4x \,
( i M \bar \psi_i T^a \omega^a Y_i - i M \bar Y_i T^a \omega^a \psi_i)
\, .
\label{st9.av.2}
\end{eqnarray}
The parameters $\lambda^{(1)}_j, ~ j=1,2$, $\lambda^{(1)}_{\beta,i}, ~i=1,\dots,16$,
$\lambda_V^{(1)}$,  $\lambda_A^{(1)}$, $\rho^{(1)}_{\bar \rho}$,
$\rho^{(1)}_k, ~ k=1,2,3$ are free parameters to be fixed by a suitable choice
of normalization conditions.
The coefficients $\gamma^{(1)}_{{\cal N}_j(x)}$, $j=1,\dots,6$
are given by vacuum diagrams associated to 
superficially convergent Feynman amplitudes 
(and their derivatives w.r.t external momenta) evaluated at zero momentum. 

For $M=0$ the functional $t^4 \GG^{(1)}_m$ becomes ${\cal S}_0$-invariant.
For $M \neq 0$ it is not ${\cal S}_0$-invariant since it 
contains a non-symmetric contribution 
to the gauge boson-fermion-fermion vertex and to the fermionic antifield couplings.
These non-symmetric terms are required in order to fulfill
the one-loop ST identities in eq.(\ref{exp.4}) 
for the complete functional $\GG^{(1)}_m$. 
They are generated by the presence of cohomologically non-trivial mass
parameters in the model.

\medskip
We notice that with the choice $\rho_{\bar \rho}^{(1)}=0$
all terms entering in eq.(\ref{st9.av.2}) are IR safe, fulfilling
the power-counting criteria of \cite{Bogolyubov:nc}-\cite{Lowenstein:1975rg}.
Therefore IR convergence in the massless limit $m \rightarrow 0$
is guaranteed provided that the dependence of $\GG^{(1)}_m$
on $\bar \rho$ only happens via the antifields in eq.(\ref{exp.6}).

\section{Consistency conditions}\label{sec:WZ}

We look back at eqs.(\ref{st23.1}) and (\ref{st23}).
By eliminating the $\rho$'s between the equations we find a set
of consistency conditions (i.e. relations between superficially
comvergent Feynman amplitudes that must be automatically fulfilled
in any regularization scheme compatible with the Quantum Action Principle):
\begin{eqnarray}
&& 2 \gamma^{(1)}_{{\cal N}_6(x)} = \gamma^{(1)}_{{\cal N}_4(x)} + \gamma^{(1)}_{{\cal N}_5(x)} \,\, .
\label{cons.2}
\end{eqnarray}
The origin of these consisitency conditions is the
Wess-Zumino consistency condition valid for the
ST breaking $\Delta^{(1)}={\cal S}_0(\G_m^{(1)})$:
\begin{eqnarray}
{\cal S}_0 (\Delta^{(1)})=0 \, .
\label{cons.3}
\end{eqnarray}
$\G_m^{(1)}$ is any generally  non-symmetric 
vertex functional obtained by
a regularization procedure consistent with
the Quantum Action Principle and including a generic set of
finite counterterms compatible with power-counting.
As is clear, eq.(\ref{cons.3}) is a direct consequence of the
nilpotency of ${\cal S}_0$.

If one does not impose eq.(\ref{st.comm.1}) but keep
$\rho_4^{(1)}, \rho_5^{(1)}, \rho_6^{(1)}$ arbitrary 
the ST identities are broken.  
%$\left . \GG_m^{(1)} \right |_{\rho^{(1)}_{4,5,6}}$.
The breaking term %can be explicitly computed and 
is given by
\begin{eqnarray}
\Delta^{(1)} & = & \int d^4x \, \Big (
a^{(1)} 
\bar \psi_i \slsh{\partial} \omega^a T^a \psi_i   
+ b^{(1)} 
\bar \psi_i T^a f^{abc} \slsh{A}^b \omega^c \psi_i 
\nonumber \\
&& + e^{(1)}
i \bar\psi_i T^a f^{abc} \omega^b \omega^c Y_i 
+ \bar e^{(1)}
i \bar Y_i T^a f^{abc} \omega^b \omega^c \psi_i 
\Big )  \, .
\label{cons}
\end{eqnarray}
The explicit values of the coefficients $a^{(1)}, b^{(1)}, e^{(1)}$
and $\bar e^{(1)}$ are 
\begin{eqnarray}
&& a^{(1)} = M [  (\rho_5^{(1)} - \rho_4^{(1)}) - \rho_6^{(1)}
+ i (\gamma^{(1)}_{{\cal N}_1(x)} + \gamma^{(1)}_{{\cal N}_2(x)}
     - \gamma^{(1)}_{{\cal N}_3(x)} ) ] \, , \nonumber \\
&& b^{(1)} = -M \rho_6^{(1)} + M \gamma^{(1)}_{{\cal N}_5(x)} + M  \gamma^{(1)}_{{\cal N}_6(x)} \, , \nonumber \\
&& e^{(1)} = - \frac{M}{2} \rho_6^{(1)} + M \gamma^{(1)}_{{\cal N}_6(x)} \, ,
\nonumber \\
&& \bar e^{(1)} = - \frac{M}{2} \rho_6^{(1)} + M \gamma^{(1)}_{{\cal N}_6(x)} \, .
\label{coeff}
\end{eqnarray}
We can now compute the ${\cal S}_0$-variation of $\Delta^{(1)}$:
\begin{eqnarray}
{\cal S}_0 (\Delta^{(1)}) & = & \int d^4x \,  \Big ( (e^{(1)} - \bar e^{(1)})
(-i M \bar \psi_i T^a f^{abc} \omega^b \omega^c \psi_i) \nonumber \\
& &  ~~~~~~~
+ (e^{(1)} - \bar e^{(1)}) (-\bar \psi_i T^a f^{abc} \omega^b \omega^c \slsh{\partial} \psi_i) \nonumber \\
& &  ~~~~~~~
+ e^{(1)} (i \bar \psi_i f^{abc} \omega^b \omega^c T^a T^d \slsh{A}^d \psi_i)
+ \bar e^{(1)} (-i \bar \psi_i \slsh{A}^d T^d T^a f^{abc} \omega^b \omega^c \psi_i)
\nonumber \\
& & ~~~~~~ + (2 \bar e^{(1)} - b^{(1)}) \bar \psi_i f^{abc} \slsh{\partial} \omega^b \omega^c T^a \psi_i \nonumber \\
& & ~~~~~~ -b^{(1)} \bar \psi_i \frac{T^a}{2} f^{adj}f^{jbc} \slsh{A}^d \omega^b \omega^c \psi_i \Big ) \, .
\label{cons.6.5}
\end{eqnarray}
The above expression has to be zero according to eq.(\ref{cons.3}).
The first and second line then yield
\begin{eqnarray}
e^{(1)} = \bar e^{(1)}  \, .
\label{cons.6.6}
\end{eqnarray}
The fourth line gives 
\begin{eqnarray}
b^{(1)} = 2 \bar e^{(1)} \, .
\label{cons.6.7}
\end{eqnarray}
By using eq.(\ref{cons.6.6}) into eq.(\ref{cons.6.5}) 
eq.(\ref{cons.6.5}) becomes after some algebra
\begin{eqnarray}
{\cal S}_0(\Delta^{(1)}) & = & 
\int d^4x \,  \Big ( 
 (2 \bar e^{(1)} - b^{(1)}) \bar \psi_i f^{abc} \slsh{\partial} \omega^b \omega^c T^a \psi_i \nonumber \\
& & ~~~~~~ +
(e^{(1)} - \frac{b^{(1)}}{2}) \bar \psi_i \frac{T^a}{2} f^{adj}f^{jbc} \slsh{A}^d \omega^b \omega^c \psi_i \Big ) \, .
\label{cons.6.8}
\end{eqnarray}
This yields in addition to eqs.(\ref{cons.6.6})-(\ref{cons.6.7})
\begin{eqnarray}
e^{(1)} = \frac{b^{(1)}}{2} \, ,
\label{cons.6.9}
\end{eqnarray}
which is indeed a consequence of eqs.(\ref{cons.6.6})-(\ref{cons.6.7}).
The condition in eq.(\ref{cons.6.6}) guarantees the equality of
the coefficients of the monomials $\int d^4x \, i \bar\psi_i T^a f^{abc} \omega^b \omega^c Y_i$ and its bar-conjugated
$\int d^4x \, i \bar Y_i T^a f^{abc} \omega^b \omega^c \psi_i$, 
which is verified by
the expressions in eq.(\ref{coeff}). On the other hand
eq.(\ref{cons.6.7}) gives
\begin{eqnarray}
&& 0 = -M \rho_6^{(1)} + M \gamma^{(1)}_{{\cal N}_4(x)} + M  \gamma^{(1)}_{{\cal N}_5(x)}
- 2 ( - \frac{M}{2} \rho_6^{(1)} + M \gamma^{(1)}_{{\cal N}_6(x)} ) \nonumber \\
&& = M \gamma^{(1)}_{{\cal N}_4(x)} + M  \gamma^{(1)}_{{\cal N}_5(x)}
   - 2 M \gamma^{(1)}_{{\cal N}_6(x)} \, ,
\label{cons.6.10}
\end{eqnarray}
i.e. the consistency condition in eq.(\ref{cons.2}).

\section{Higher orders}\label{sec:higher}

Higher order ST identities involve the solution
of an inhomogeneous equation for $t^4 \GG^{(n)}_m$, $n >1$:
\begin{eqnarray}
{\cal S}_0(t^4 \GG^{(n)}_m) = -{\cal S}_0((1-t^4) \GG^{(n)}_m)
- \sum_{j=1}^{n-1} (\GG^{(j)},\GG^{(n-j)}) \, .
\label{higher.1}
\end{eqnarray}
The parenthesis in the above equation is defined by
\begin{eqnarray}
(X,Y) = \int d^4x \, \Big (
\frac{\delta X}{\delta \hat A_\mu^{a*}} \frac{\delta Y}{\delta A^{a\mu}}
+
\frac{\delta X}{\delta \hat \omega^{a*}} \frac{\delta Y}{\delta \omega^a}
- \frac{\delta X}{\delta Y_i} 
  \frac{\delta Y}{\delta \bar \psi_i} 
+ \frac{\delta X}{\delta \bar Y_i} 
  \frac{\delta Y}{\delta \psi_i} 
\Big ) \, .
\label{bracket}
\end{eqnarray}
The R.H.S. of the eq.(\ref{higher.1}) is known (it contains 
superficially convergent $n$-order Feynman amplitudes
and known lower dimensional contributions coming from
the bracket $(\GG^{(j)},\GG^{(n-j)})$).
Again we can limit ourselves to the projection of
eq.(\ref{higher.1}) on the subspace of dimension $\leq 5$
in the fields, antifields, external sources and 
their derivatives of ghost number one,
since by the QAP higher dimensional terms are absent
(under the recursive assumption that the ST identities have been
restored up to order $n-1$).

Since ${\cal S}_0$ is nilpotent the ${\cal S}_0$-variation of
 eq.(\ref{higher.1}) gives
\begin{eqnarray}
{\cal S}_0 ( \sum_{j=1}^{n-1} (\GG^{(j)},\GG^{(n-j)}) ) =0 \, .
\label{higher.2}
\end{eqnarray}
In the absence of anomalies by eq.(\ref{higher.2}) there exists a functional
$\Theta^{(n)}$ such that
\begin{eqnarray}
\sum_{j=1}^{n-1} (\GG^{(j)},\GG^{(n-j)}) = {\cal S}_0 (\Theta^{(n)}) \, .
\label{higher.3}
\end{eqnarray}
This functional is not unique (one can add any ${\cal S}_0$-invariant
with ghost number zero
to $\Theta^{(n)}$ without violating eq.(\ref{higher.3})).
The knowledge of any particular solution to eq.(\ref{higher.3}) is
enough for the present purposes. Since the L.H.S. of eq.(\ref{higher.3})
 is completely fixed in terms of lower order Feynman amplitudes,
the determination of $\Theta^{(n)}$ is a purely algebraic problem.

We insert eq.(\ref{higher.3}) in eq.(\ref{higher.1}) and
obtain a homogeneous equation for $\GG_m^{(n)} + \Theta^{(n)}$:
\begin{eqnarray}
{\cal S}_0 ( \GG^{(n)}_m + \Theta^{(n)}) = 0 \, .
\label{higher.4}
\end{eqnarray}
At this stage the technique developed in Sect.~\ref{sec:oneloop} can be applied.
We obtain
\begin{eqnarray}
{\cal S}_0 ( t^4 \GG^{(n)}_m) = 
- {\cal S}_0 ( (1-t^4) \GG^{(n)}_m) - {\cal S}_0 (\Theta^{(n)}) \, .
\label{higher.5}
\end{eqnarray}
The R.H.S. is known. The contribution from the bracket in eq.(\ref{higher.1}),
which controls the non-linearity of higher-order ST identities, is 
taken into account by the last term in the R.H.S. of eq.(\ref{higher.5}).

\medskip
The explicit determination of the functional $\Theta^{(n)}$ 
for arbitrary values of the free parameters $\lambda^{(j)},\rho^{(j)}$,
$j<n$ may be a difficult task and results in very complicated expressions.
Examples for the pure Yang-Mills sector are given in \cite{Quadri:2003ui}.

However the problem simplifies considerably if the freedom in the choice
of $\lambda^{(j)},\rho^{(j)}$ is used in order to achieve
the simplest possible form of the contribution generated by the bracket 
in eq.(\ref{higher.1}).

Let us illustrate this point on the model at hand.
We start from two-loop order.
It turns out that in the present model there is a natural choice of
the normalization conditions such that
\begin{eqnarray}
(\GG^{(1)},\GG^{(1)}) = 0 \,  ~~~~~ {\mbox{up to dimension 5}}.
\label{h2}
\end{eqnarray}
The proof is as follows. Let us choose 
\begin{eqnarray}
&& \lambda_1^{(1)}=0 \, , ~~~ \lambda_2^{(1)} = 0 \, , ~~~
   \rho_1^{(1)} = 0 \, , ~~~ \rho_{\bar \rho}^{(1)} = 0 \, .
\label{h3}
\end{eqnarray}
Moreover we choose $\rho_2^{(1)}$ in order to cancel the
$\bar Y_i, Y_i$-terms in $t^4 \GG^{(1)}$. This can be done since
\begin{eqnarray}
{\cal S}_0 (\int d^4x \, \hat \omega^{a*} \omega^a)
 & = & \delta (\int d^4x \, \hat \omega^{a*} \omega^a) \nonumber \\
& = &
\int d^4x \, \Big ( - \omega^a \frac{\delta \G^{(0)}}{\delta \omega^a} 
+ \hat \omega^{a*} (-\frac{1}{2} f^{abc} \omega^b \omega^c) \Big ) \nonumber \\
& = & \int d^4x \, \Big ( - \hat A^{*a}_\mu (D^\mu \omega)^a 
- \hat \omega^{a*} (-\frac{1}{2} f^{abc} \omega^b \omega^c) \nonumber \\
& & ~~~~~~~~~
+ i \bar \psi_i T^a \omega^a Y_i
- i \bar Y_i \omega^a T^a \psi_i \Big ) \, .
\label{h4}
\end{eqnarray}
By choosing 
\begin{eqnarray}
\rho_2^{(1)} =  - 2 M \gamma^{(1)}_{{\cal N}_6}
\label{h4.bis}
\end{eqnarray}
we obtain
the desired condition on $t^4 \GG^{(1)}$.

Finally we choose $\rho_3^{(1)}$ in such a way to 
set the coefficient of the monomial $\int d^4x \, \bar \psi_i \slsh{A}^a T^a \psi_i$
equal to zero, namely
\begin{eqnarray}
\rho_3^{(1)}  
= - i M
(\gamma^{(1)}_{{\cal N}_1(x)} + \gamma^{(1)}_{{\cal N}_2(x)} - \gamma^{(1)}_{{\cal N}_3(x)} )
+2 M \gamma^{(1)}_{{\cal N}_6(x)} \, .
\label{h5}
\end{eqnarray}
The resulting expression for $t^4 \GG^{(1)}$ is
\begin{eqnarray}
t^4 \GG^{(1)} & = & 
+ \sum_{i=1}^{16} \lambda^{(1)}_{\beta,i} \Lambda_{\beta,i}(x) 
 + \lambda_V^{(1)} \int d^4x \, \beta^V_\mu \bar \psi \gamma^\mu \psi 
 + \lambda_A^{(1)} \int d^4x \, \beta^A_\mu \bar \psi \gamma^\mu \gamma^5 \psi 
\nonumber \\ 
& & 
+ \rho_2^{(1)} \left . \delta \int d^4x \, (\hat \omega^{a*} \omega^a) \right |_{Y_i = \bar Y_i=0} %\nonumber \\
%& & 
+ \rho_3^{(1)} \int d^4x \, i \bar \psi_i \slsh{\partial} \psi_i \, .
\label{h6}
\end{eqnarray}
By eq.(\ref{h6})
\begin{eqnarray}
( t^4 \GG^{(1)}, t^4 \GG^{(1)}) = (\rho_2^{(1)})^2 
( \left . \delta \int d^4x \, (\hat \omega^{a*} \omega^a) \right |_{Y_i = \bar
    Y_i=0}, \left . \delta \int d^4x \, (\hat \omega^{a*} \omega^a) \right |_{Y_i = \bar Y_i=0}) \, . \nonumber \\
\label{h7}
\end{eqnarray}
By power-counting, since $\lambda^{(1)}_2$ has been chosen equal to zero,
$( t^4 \GG^{(1)}, t^4 \GG^{(1)})$ coincides up to dimension $5$
with $(\GG^{(1)}, \GG^{(1)})$. Therefore we can limit ourselves
to the study of eq.(\ref{h7}).

The R.H.S. of eq.(\ref{h7}) is zero. This can be easily seen
by noticing that from eq.(\ref{h4})
$\left . \delta \int d^4x \, (\hat \omega^{a*} \omega^a) \right |_{Y_i = \bar Y_i=0}$ is, up to an overall minus sign, the antifield-dependent part 
of the classical action of pure Yang-Mills theory. 
Eq.(\ref{h7}) is nothing but the classical 
ST identities for $\left . \delta \int d^4x \, (\hat \omega^{a*} \omega^a) \right |_{Y_i = \bar Y_i=0}$.  Since the latter is BRST invariant, eq.(\ref{h7}) vanishes.
Therefore there exists a choice of normalization conditions given 
by eqs.(\ref{h3}),(\ref{h4.bis}) and (\ref{h5})
allowing to discard the inhomogeneous contribution originated by the bracket
$(\GG^{(1)},\GG^{(1)})$ in the relevant sector of dimension $\leq 5$.
With this special choice of normalization conditions the two-loop
ST identities read
\begin{eqnarray}
{\cal S}_0 ( \GG^{(2)} ) = 0 \, .
\label{2loop.1}
\end{eqnarray}
The discussion of eq.(\ref{2loop.1}) proceeds in the same way as
for eq.(\ref{exp.4}).
The argument can be recursively repeated to all orders in the loop
expansion.
If the following normalization conditions
are imposed order by order in perturbation theory for
$j=1,2,\dots,n-1$
\begin{eqnarray}
&& 
\!\!\!\!\!\!\!\!\!\!\!\!\!\!\!\!\!\!\!\!
\lambda_1^{(j)}=0 \, , ~~~ \lambda_2^{(j)} = 0 \, , ~~~
   \rho_1^{(j)} = 0 \, , ~~~ \rho_{\bar \rho}^{(j)} = 0 \, , \nonumber \\
&& 
\!\!\!\!\!\!\!\!\!\!\!\!\!\!\!\!\!\!\!\!
\rho_2^{(j)} =  - 2 M \gamma^{(j)}_{{\cal N}_6} \, , ~~~
\rho_3^{(j)}  
= - i M
(\gamma^{(j)}_{{\cal N}_1(x)} + \gamma^{(j)}_{{\cal N}_2(x)} - \gamma^{(j)}_{{\cal N}_3(x)} )
+2 M \gamma^{(j)}_{{\cal N}_6(x)} 
\label{h_allorder}
\end{eqnarray}
the action-like part of the $n$-th order symmetric vertex functional
is
\begin{eqnarray}
t^4 \GG^{(n)} & = &
\lambda^{(n)}_1 \int d^4x \, F_{\mu\nu}^a F^{\mu\nu a}
                   +\lambda^{(n)}_2 \int d^4x \, \bar \psi_i \psi_i
\nonumber \\
&   & + \sum_{i=1}^{16} \lambda^{(n)}_{\beta,i} \Lambda_{\beta,i}(x)
      + \lambda_V^{(n)}  \int d^4x \, \beta^V_\mu \bar \psi
                   \gamma^\mu \psi 
      + \lambda_A^{(n)}  \int d^4x \, \beta^A_\mu \bar \psi
                   \gamma^\mu \gamma^5 \psi 
\nonumber \\
&   & +\rho^{(n)}_{\bar \rho} \Big ( \frac{m^2 }{2} \int d^4x \, (A_\mu^a)^2 - 
\int d^4x \, \bar \rho A_\mu^a \partial^\mu \omega^a \Big ) \nonumber \\
&   & +\rho^{(n)}_1 \delta \int d^4x \, (\hat A^{a*}_\mu A^{a\mu})
      +\rho^{(n)}_2 \delta \int d^4x \, (\hat \omega^{a*} \omega^a )
\nonumber \\
&   &  +\rho^{(n)}_3 \int d^4x \, i \bar \psi_i \slsh{\partial} \psi_i 
\nonumber \\
&  & + (\rho_3^{(n)}  
+ i M
(\gamma^{(n)}_{{\cal N}_1(x)} + \gamma^{(n)}_{{\cal N}_2(x)} - \gamma^{(n)}_{{\cal N}_3(x)} )
-2 M \gamma^{(n)}_{{\cal N}_6(x)} )
 \int d^4x \, \bar \psi_i \slsh{A}^a T^a \psi_i 
 \nonumber \\
&  & + 2 \gamma^{(n)}_{{\cal N}_6} \int d^4x \,
( i M \bar \psi_i T^a \omega^a Y_i - i M \bar Y_i T^a \omega^a \psi_i)
\, .
\label{st9.av.3}
\end{eqnarray}
Moreover, with the choice $\rho_{\bar \rho}^{(j)}=0$, $j=1,\dots,n$
all terms entering in eq.(\ref{st9.av.2}) are IR safe, fulfilling
the power-counting criteria of \cite{Bogolyubov:nc}-\cite{Lowenstein:1975rg}.
This in turn implies that the massless limit $m \rightarrow 0$
($s \rightarrow 1$) of $\GG_m$ exist:
\begin{eqnarray}
\GG = \lim_{m \rightarrow 0} \GG_m
\label{lim.1}
\end{eqnarray}
and obeys the following ST identities of Yang-Mills theory
with massive fermions 
\begin{eqnarray}
{\cal S}(\GG) & = & \int d^4x \, \Big (
\frac{\delta \GG}{\delta A_\mu^{a*}} \frac{\delta \GG}{\delta A^{a\mu}}
+
\frac{\delta \GG}{\delta \omega^{a*}} \frac{\delta \GG}{\delta \omega^a}
+
B^a \frac{\delta \GG}{\delta \bar \omega^a}
\nonumber \\
& & ~~~~~~~~~
- \frac{\delta \GG}{\delta Y_i} 
  \frac{\delta \GG}{\delta \bar \psi_i} 
+ \frac{\delta \GG}{\delta \bar Y_i} 
  \frac{\delta \GG}{\delta \psi_i} 
\Big ) = 0 \, .
\label{st.final}
\end{eqnarray}

\section{Conclusions}\label{sec:concl}

By applying the method of the ST parameterization of the symmetric 
(fulfilling the ST identities) vertex functional,
a complete characterization of the action-like part of the symmetric
quantum effective action has been achieved for Yang-Mills theory
with massive fermions in the presence of singlet axial-vector
currents.

An IR regulator $m$ has been introduced within the BPHZL scheme
and the relevant extended ST identities, parameterizing the soft-breaking
induced by the intermediate gauge bosons and ghost masses, have been analyzed. 
The IR regulated symmetric vertex functional $\GG_m$ has been constructed.
The limit $\GG_m$ for $m\rightarrow 0$ is seen to exist since the 
BPHZL IR power-counting criteria hold true at any order in the loop expansion.
In this limit the extended ST identities reduce to those of Yang-Mills
theory with massive fermions in eq.(\ref{st.final}).

\medskip
The ST parameterization reveals an interesting 
geometric structure also for the finite
action-like part of the symmetric quantum effective action. 

Since a cohomologically non-trivial mass parameter like the fermion mass $M$
enters into the theory, the action-like functional $t^4\GG_m$ is not
invariant under the linearized classical ST operator ${\cal S}_0$.
The non-invariant action-like terms 
proportional to $M$ are needed in order to fulfill the ST
identities for the whole vertex functional $\GG_m$.

Moreover, we have shown that a natural set of normalization conditions
exists allowing to reduce higher-order ST identities to a linear
homogeneous problem. 
This is a rather non-trivial consequence of the geometric structure
of $t^4 \GG_m$ unveiled by the ST parameterization of the quantum 
effective action.

By exploiting this set of normalization
conditions the expression of the action-like part of the symmetric
vertex functional to any order in the loop expansion has been given.

\medskip

We remark that the inclusion of composite operators like
the singlet axial-vector currents does not pose any additional problem
in the application of the ST parameterization technique.
Hence the latter also provides an efficient way to treat BRST-invariant 
composite operators in those theories where no invariant regularization
scheme is known. 

\medskip

We would like finally to comment on the use of the BPHZL regularization scheme
in order to construct the symmetric $n$-th order quantum effective action.
In this scheme an IR regulator is introduced in a way consistent with
power-counting (so that each factor of $m = \mu (s-1)$ counts as one).

It should be noticed that the algebraic structure underlying the use 
of the IR regulator is independent of the particular regularization
scheme adopted and  can be thus translated without difficulty  
to other regularization procedures, 
like for instance  Dimensional Regularization (DR)  
(in the formulation compatible with  the Quantum Action Principle 
of \cite{Breitenlohner:1977hr,Breitenlohner:1975hg,Breitenlohner:1976te}).
However, special care has to be paid to the subtraction chosen
to deal with the IR regulator in order to ensure that the 
power-counting properties true for the BPHZL IR regulator 
are still valid.
Provided that this requirement is fulfilled, the formulas given 
in eqs.(\ref{st9.av.2}) and  (\ref{st9.av.3}) can be regarded as
 regularization scheme-independent results.

\section*{Acknowledgments}

Useful discussions with R.~Ferrari and D.~Maison are gratefully
acknowledged. The Author also would like to thank the warm
hospitality of the Theory Group at Max-Planck-Institut f\"ur Physik
(Werner-Heisenberg-Institut), where part of this work was completed.

\appendix

\section{BRST symmetry}\label{app:brst}

The BRST symmetry of the model is
\begin{eqnarray}
&& s A^a_\mu = (D_\mu \omega)^a \equiv \partial_\mu \omega^a + f^{abc} A^b_\mu \omega^c \, , ~~~~~
s \omega^a = -\frac{1}{2} f^{abc} \omega^b \omega^c \, , 
\nonumber \\
&& s \psi_i = i \omega^a T^a \psi_i \, , ~~~~~
   s \bar \psi_i =  i \bar \psi_i T^a \omega^a \, .
\label{l2}
\end{eqnarray}
The following conjugation rules hold:
\begin{eqnarray}
s \bar b = \overline{s b} \, , ~~~~ s \bar f = - \overline{s f} \, ,
\label{l3}
\end{eqnarray}
with $b$ boson and $f$ fermion.

\medskip
The introduction of the BPHZL mass regulator $m$ in eq.(\ref{ir.3}) requires
an additional BRST pair $(\bar \rho,m^2)$ 
\begin{eqnarray}
s \bar \rho = m^2 \, , ~~~~ s m^2 = 0 \, ,
\label{app.1}
\end{eqnarray}
with $\bar \rho$ an anticommuting constant parameter.

\medskip
The IR regularized BRST-invariant classical action is
\begin{eqnarray}
\G^{(0)}_m & = & \int d^4x \, \Big ( -\frac{1}{4g^2} F^a_{\mu\nu} F^{\mu\nu a} 
          + i \bar \psi_i \slsh{D} \psi_i - M \bar \psi_i \psi_i 
+ \beta^V_\mu \bar \psi \gamma^\mu \psi + 
                    \beta^A_\mu \bar \psi \gamma^\mu \gamma^5 \psi 
\nonumber \\
& & 
~~~ + \frac{1}{2} m^2 (A_\mu^a)^2 + m^2 \bar \omega^a
\omega^a  
- \bar \rho A_\mu^a \partial^\mu \omega^a - \bar \rho B^a \omega^a 
- \frac{1}{2} \bar \rho \bar \omega^a f^{abc} \omega^b \omega^c \nonumber \\
& & ~~~
+ \alpha \frac{(B^a)^2}{2} - B^a \partial A^a + 
\bar \omega^a \partial^\mu ( D_\mu \omega)^a 
\nonumber \\
& & ~~~ + A_\mu^{a*} (D^\mu \omega)^a
+ \omega^{a*} (-\frac{1}{2} f^{abc} \omega^b \omega^c ) 
- i \bar \psi_i T^a \omega^a Y_i
+ i \bar Y_i \omega^a T^a \psi_i \Big ) \nonumber \\
\label{class.act}
\end{eqnarray}

\medskip
The UV dimension of the fields, antifields and external sources
is assigned according to the following table:
\begin{center}
\noindent
\begin{tabular}{|c|c|c|c|c|c|c|c|c|c|c|c|c|}
\hline
Field & $A_\mu^a$ & $\omega^a$ & $\bar \omega^a$ & $B^a$ & $\psi_i$ & $\bar
\psi_i$ & $A_\mu^{a*}$ & $\omega_a^*$ & $\bar Y_i$ & $Y_i$ & $\beta_\mu^V$ &
$\beta_\mu^A$ \cr
\hline
UV dim. & 1 & 1 & 1 & 2 & 3/2 & 3/2 & 2 & 2 & 3/2 & 3/2 & 1 & 1 \cr
\hline
\end{tabular}
\vskip 0.3 truecm
{\bf Table 1} - UV dimension of the fields, antifields and
external sources
\end{center}

\medskip
$\bar \psi,\psi$ have IR dimension two. The IR dimension of the other
fields coincides with the UV dimension.
Under parity $\beta_\mu^V$ is even and $\beta_\mu^A$ odd.

\section{Decomposition of $\delta$}\label{app:decomp}

The decomposition of $\delta$ according to the UV dimension
of the fields, external sources and their derivatives is
\begin{eqnarray}
\begin{array}{ll}
\delta_0 A^a_\mu = 0 \, , ~~~~~ & \delta_1 A^a_\mu = (D_\mu \omega)^a \, , 
\nonumber \cr
\delta_0 \omega^a = 0 \, , ~~~~~ & \delta_1 \omega^a = -\frac{1}{2} f^{abc} \omega_b \omega_c \, , \nonumber \cr
\delta_0 \bar \omega_a = 0 \, , ~~~~~ & \delta_1 \bar\omega_a = B_a \, , \nonumber \cr
\delta_0 B^a = 0 \, ,~~~~~ & \delta_1 B^a = 0 \, , \nonumber \cr
\delta_0 \psi_i = 0 \, , ~~~~~ & \delta_1 \psi_i = i \omega^a T^a \psi_i \, ,
\nonumber \\
\delta_0 \bar \psi_i = 0 \, , ~~~~~ & \delta_1 \bar \psi_i = i \bar \psi_i T^a \omega_a \, , \nonumber \\
\delta_0 \hat A^{a*}_\mu = 0 \, , ~~~~~ & \delta_1 \hat A^{a*}_\mu =
\left . \frac{\delta \G^{(0)}}{\delta A^a_\mu} \right |_{B_a=0,m^2=0} \, ,
\nonumber \\
\delta_0 \hat \omega^{a*} = 0 \, , ~~~~~ & \delta_1 \hat \omega^{a*} = 
\left . \frac{\delta \G^{(0)}}{\delta \omega_a} \right |_{B^a=0,m^2=0}\, , \nonumber \\
\delta_0 \bar Y_i = M \bar \psi_i \, , ~~~~~ & \delta_1 \bar Y_i = 
\left . \frac{\delta \G^{(0)}}{\delta \psi_i} \right |_{M=0} \, ,
\nonumber \\
\delta_0 Y_i = M \psi_i \, , ~~~~~ & \delta_1 Y_i = -
\left . \frac{\delta \G^{(0)}}{\delta \bar \psi_i} \right |_{M=0} \, .
\end{array}
\label{l19}
\end{eqnarray}
\section{Invariants depending only on $\beta_V^\mu, \beta_A^\mu$
with dimension $\leq 4$}\label{app:beta_inv}
\begin{enumerate}
\item With zero derivatives:
\begin{eqnarray}
&& \Lambda_{VVVV,1} = \int d^4x \, (\beta^V)^2 (\beta^V)^2 \, , ~~~~ 
\nonumber \\
&& 
\Lambda_{AAAA,1} = \int d^4x \, (\beta^A)^2 (\beta^A)^2 \, , ~~~~
\nonumber \\
&& 
\Lambda_{AAVV,1} = \int d^4x \, (\beta^A)^2 (\beta^V)^2 \, , \nonumber \\
&& \Lambda_{AAVV,2} = \int d^4x \, (\beta^A \beta^V)^2 \, . ~~~~ 
%\nonumber \\
\label{l6.av.4}
\end{eqnarray}
\item With one derivative (if at least one $\beta_A$ is present 
we decide not to differentiate $\beta_V$):
\begin{eqnarray}
&& \Lambda_{VVV,1} = \int d^4x \, \partial \beta^V (\beta^V)^2 \, , \nonumber \\
&& \Lambda_{VVV,2} = \int d^4x \, \partial_\mu \beta^V_\nu \beta^{V \mu} \beta^{V \nu} \, , \nonumber \\
&& \Lambda_{AAV,1} = \int d^4x \, \beta^V \beta^A \partial \beta^A \, , \nonumber \\
&& \Lambda_{AAV,2} = \int d^4x \, \beta^V_\mu \beta^A_\nu \partial^\mu \beta^{A\nu} \, , \nonumber \\
&& \Lambda_{AAV,3} = \int d^4x \,\beta^V_\mu \beta^A_\nu \partial^\nu \beta^{A\mu} \, , \nonumber \\
&& 
\Lambda_{AVV} = \int d^4x \, \epsilon^{\mu\nu\rho\sigma}
\beta^A_\mu \beta^V_\nu \partial_\rho \beta^V_\sigma \, .
\label{l6.av.5}
\end{eqnarray}
\item With two derivatives:
\begin{eqnarray}
&& \Lambda_{AA,1} = \int d^4x \, (\partial \beta^A)^2 \, , \nonumber \\
&& \Lambda_{AA,2} = \int d^4x \, \partial_\nu \beta_\mu^A \partial^\nu \beta^{A\mu} \, , \nonumber \\
&& \Lambda_{AA,3} = \int d^4x \, \partial_\nu \beta_\mu^A \partial^\mu \beta^{A\nu} \, , \nonumber \\
&& \Lambda_{VV,1} = \int d^4x \, (\partial \beta^V)^2 \, , \nonumber \\
&& \Lambda_{VV,2} = \int d^4x \, \partial_\nu \beta_\mu^V \partial^\nu \beta^{V\mu} \, , \nonumber \\
&& \Lambda_{VV,3} = \int d^4x \, \partial_\nu \beta_\mu^V \partial^\mu \beta^{V\nu} \, .
\label{l6.av.6}
\end{eqnarray}
\end{enumerate}
We denote them by $\int d^4x \, \Lambda_{\beta,i}(x)$,
$i=1,\dots,16$.

\section{Monomials of dimension $\geq 5$ contributing to the
ST breaking terms}\label{app:brkg_list}

We list the monomials of dimension $\geq 5$ contributing to
the ST breaking terms.
We use the freedom to perform an integration by parts in order 
not to differentiate $\bar \psi_i$ in the ST breaking terms.
\begin{eqnarray}
&& {\cal N}_1(x) = i \bar Y_i \slsh{\partial} \omega^a T^a \psi_i \, ,
\nonumber \\
&& {\cal N}_2(x) = i \bar \psi_i T^a \slsh{\partial} \omega^a Y_i \, ,
\nonumber \\
&& {\cal N}_3(x) = i \slsh{\partial} \bar \psi_i T^a \omega^a Y_i \, ,
\nonumber \\
&& {\cal N}_4(x) = \bar Y_i T^a f^{abc} \slsh{A}^b \omega^c \psi_i \, ,
\nonumber  \\
&& {\cal N}_5(x) = \bar \psi_i T^a f^{abc} \slsh{A}^b \omega^c Y_i \, ,
\nonumber \\
&& {\cal N}_6(x) = i \bar Y_i T^a f^{abc} \omega^b \omega^c Y_i \, .
\label{st20}
\end{eqnarray}
%

%\section{Decoupling of the IR regulator}\label{app:restr}
%

\end{document}